\documentclass[aps,pre, preprint,groupedaddress]{revtex4-1}
\usepackage{times}
\usepackage{graphicx}
\usepackage{psfrag}
\usepackage{ae}
\usepackage{amsmath,amssymb}

\begin{document}

\title{Flow and structure of fluids in functionalized nanopores}
 
\author{Jos\'e Rafael Bordin} 
\email{josebordin@unipampa.edu.br}
\affiliation{Campus Ca\c capava do Sul, Universidade Federal
do Pampa, Av. Pedro Anuncia\c c\~ao, 111, CEP 96570-000, 
Ca\c capava do Sul, RS, Brazil}

\author{Marcia C. Barbosa} 
\email{marcia.barbosa@ufrgs.br}
\affiliation{Instituto 
de F\'{\i}sica, Universidade Federal
do Rio Grande do Sul, Caixa Postal 15051, CEP 91501-970, 
Porto Alegre, RS, Brazil}

\begin{abstract}
  We investigate through non-equilibrium Molecular Dynamics simulations
  the structure and flow of fluids in functionalized 
  nanopores. The nanopores are modeled as cylindrical structures with
  solvophilic and solvophobic sites. Two fluids are modeled. The first is a standard
  Lennard Jones fluid. The second one is modeled with a isotropic
  two-length scale potential, which exhibits in bulk water-like anomalies.
  Our results indicates distinct dependence of the overall mass flux 
  for each species of fluid  with 
  the number of solvophilic sites for different nanotubes radii. Also,
  the density and fluid structure are dependent from the nanotube radius
  and the solvophilic properties of the nanotube. This indicates that
  the presence of a second length scale in the fluid-fluid interaction
  will lead to distinct behavior. Also, our results shows that
  chemically functionalized nanotubes with different radius
  will have distinct nano-fluidic features. Our results are explained
  in the basis of the characteristic scale fluid properties and the effects
  of nanoconfinement.
\end{abstract}


\maketitle

\section{Introduction}

Most  
liquids  contract upon
cooling at constant pressure and diffuse slower upon 
compression. This is not the case of a wide range
of materials that expand as the temperature is 
decreased and move faster as the pressure grows. 
Liquid water is the most known  of 
this anomalous materials~\cite{URL} but it is not the only one. 
The  maximum in the diffusion coefficient at constant temperature
was observed in  water~\cite{Ne02a}but also 
in  silicon~\cite{Mo05} and 
silica~\cite{Sa03}.
The maximum in the density is present in water~\cite{Ke75} but also
in silicon~\cite{Sa03}, silica~\cite{Sh06},
Te~\cite{Th76}, Bi~\cite{Handbook}, 
Si~\cite{Ke83}, $Ge_{15}Te_{85}$~\cite{Ts91},  liquid 
metals~\cite{Cu81}, graphite~\cite{To97} and 
$BeF_2$~\cite{An00}.

The origin of the unusual behavior observed in 
anomalous materials is the presence of 
two characteristics length scales. While non anomalous
liquids can be described on the framework of the van
der Waals one length scale potential, anomalous
materials exhibit two characteristic scales.
Then it became natural to associate
the thermodynamic and dynamic anomalous behavior
of these materials with a core-softened (CS) potentials with two 
length scales (TLS)
in the bulk~\cite{Ma05,Ja98,Sc00a,Xu05, Oliveira06a, Fomin11} or
under 
confinement~\cite{Krott13a,Krott13b,Krott14a,Bordin14a, BoK14c, Krott15a, Krott15b}.

Recently a new puzzle was added to list of anomalous behaviors of water.
Experiments and simulations show that the  mobility of water  through nanotubes membranes 
 exceeds the values calculated from continuum hydrodynamics models by
more than three orders of 
magnitude~\cite{Holt06,Qin11, thomas09, Roy14, Bordin12b, Bordin13a, Bordin14b}. This
behavior is also observed
in non anomalous materials such as  ionic 
liquids ~\cite{Chaban14}, hydrogen, methane, nitrogen, 
air, of oxygen, of argon~\cite{Majumder11, Yasuoka15}.
In this case, however,   the 
fast flow exceeds predictions 
 by only one order of magnitude~\cite{Holt06,Qin11}. Naturally,
two questions arise: what is the mechanism behind the 
fast flow in nanoconfinement and why in the case of water
it is much faster.

In the particular case of water, the hydrophobicity
adds up to this already complex 
problem~\cite{Mosko14,giovambattista09b}. While 
for the
hydrophobic wall-water interaction in an homogeneous 
tube the flow is faster than in the pure hydrophilic wall-water
interaction~\cite{Li12,Martens08}, for an
heterogeneous wall-water system the  
flow is faster as the system
becomes fully  hydrophilic~\cite{Mosko14,Hummer01}.

The existence of confining structures in which 
hydrophobic and hydrophilic sites are present
is not  just a theoretical assumption.
Recent methods allowed the synthesis of nanotubes
similar to CNTs, as Boron-Nitride nanotubed (BNNTs)~\cite{Chopra95} and 
carbon doped BNNTs~\cite{Wei10}.
Chemically functionalized nanotubes can have hydrophobic and hydrophilic 
sites, similar
to biological channels, witch have distinct solvophobic properties depending 
on the forming
amino acids. Also, gas  adsorption and storage in doped and 
chemically treated nanotubes
have been recently investigated~\cite{Barghi16, Jin15}.
Therefore studying these systems is not only a 
theoretical challenge but it has realistic applications.

In this paper  we explore the
differences and similarities between the anomalous and non anomalous fluids
 flow by computer
simulations.
We compute  the mobility  of a water-like
 system under
solvophilic, solvophobic and mixed confinement. This
behavior is then compared with the flow of 
a non anomalous, standard LJ fluid 
also confined within an  attractive, a repulsive and a mixed walls.
Our results aim to shade some light in the controversial
results both in the water-like and non anomalous confinement in the
different types of 
wall-fluid interactions.

The paper is 
organized as follows: in Sec. II we introduce the model
and describe the methods and simulation details; the results are 
given and discussed in Sec. III; and 
in Sec. IV we present our conclusions.

\section{Model and simulational details}

In this paper all physical quantities are computed
in the standard Lennard-Jones (LJ) units~\cite{AllenTild}, as instance
\begin{equation}
\label{red1}
r^*\equiv \frac{r}{\sigma}\;,\quad \rho^{*}\equiv \rho \sigma^{3}\;, \quad 
\mbox{and}\quad t^* \equiv t\left(\frac{\epsilon}{m\sigma^2}\right)^{1/2}\;,
\end{equation}
for distance, density of particles and time , respectively,
where $\sigma$ is the distance
parameter, $\epsilon$ the energy parameter and $m$ the mass parameter.
Since all physical quantities are defined in reduced LJ units, 
the $^*$ is  omitted, in order to simplify the discussion.

\begin{figure}[ht]
 \begin{center}
 \includegraphics[clip,width=12cm]{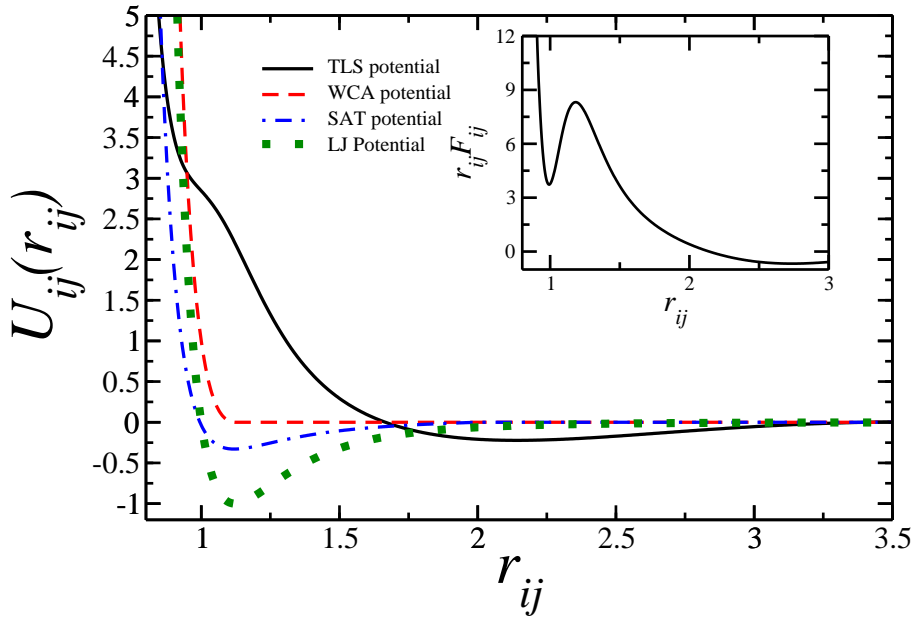}
 \end{center}
 \caption{Interactions potentials used in our simulations as function
 of particles separation $r_{ij}$. 
 The solid black line is the TLS potential used to modeled the 
anomalous fluid.
 The dashed red line is the purely repulsive WCA potential, while de 
dot--dashed blue line
 is the attractive SAT potential. The dotted green line represents the 
LJ potential. 
 Inset: $r_{ij}F(r_{ij})$ versus $r_{ij}$ for the TLS potential. According 
to Debenedetti {\it et al}~\cite{Debenedetti91}
 a potential will show anomalous behavior if there is a region 
where $(\partial r_{ij}F(r_{ij})/\partial r_{ij})> 0$.}
 \label{fig1}
 \end{figure}

  Two types of fluids are analyzed: a water-like fluid
and a non anomalous fluid. Both are modeled
by coarse-graining potentials. The water-like system is 
represented by a two length scale interaction potential
and the non anomalous  fluid is depicted 
by an one length scale interaction potential. 
In both cases the system is modeled 
by spherical particles with effective
 diameter $\sigma$ and mass $m$.
The water-like particles interact through the three
 dimensional core-softened potential (TLS):
 \begin{equation}
 U_{ij}^{\rm{TLS}}(r_{ij}) = \epsilon'\left[ \left(\frac{\sigma}{r_{ij}}\right)^{12} - 
 \left(\frac{\sigma}{r_{ij}}\right)^6 \right] + 
 \sum_{i=1}^3 \frac{B_i}{B_i^2 + (r_{ij}-C_i)^2}\;,
 \label{Potential}
 \end{equation}
 where $r_{ij}=|\vec r_i - \vec r_j|$ is the distance between fluid particles
 $i$ and $j$. The first term on the right is the standard 12-6
 LJ potential~\cite{AllenTild}, while the second term corresponds to 
 Lorentzian distributions centered at $C_i$ with amplitude $1/B_i$.
 In this paper, the potential parameter are: $k=3$, $\epsilon' = 0.6$, $B_1 = 0.3$,
 $B_2 = -1.5$, $B_3 = 2.0$, $C_1 = 1.0$, $C_2 = 1.8$ and $C_3 = 3.0$, resulting
 in the TLS potential showed in the figure~\ref{fig1}. The first length scale is 
 at $\approx 1.0$, the particle diameter. The second length scale is 
 located at $\approx 1.5$. This type of effective 
 potential exhibits the density, diffusion
and structural anomalous behavior present in water~\cite{Oliveira06a}
and it has been used as a coarse-graining model for system
with thermodynamic, dynamic and structural anomalous behavior.

The non anomalous particles  interact through the standard Lennard-Jones potential
also shown in the fig.~\ref{fig1} and given by,
\begin{equation}
\label{Potential2}
U_{{\rm {LJ}}}(r_{ij}) = 4\epsilon\left[\left(\frac{\sigma}{r_{ij}}\right)^{12} - 
 \left(\frac{\sigma}{r_{ij}}\right)^6 \right]\;,
\end{equation}
where $\epsilon$ is the depth of the attractive well. The LJ potential 
is showed
as the green dotted line in figure~\ref{fig1}. This potential
has been used to model simple one length scale system
in which the density, diffusion and structural anomalous
behavior observed in water is not seen. Here the system is analyzed
in the pressure-temperature region in which no liquid-gas phase transition
is present.

\begin{figure}[ht]
 \begin{center}
 \includegraphics[clip,width=12cm]{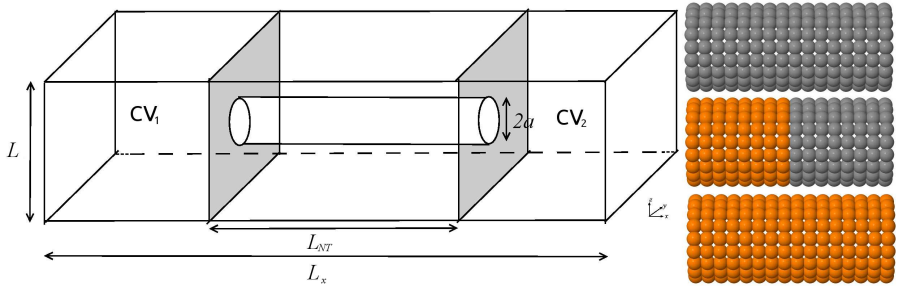}
 \end{center}
 \caption{Left: Schematic depiction of the simulation box. The system is composed of a 
high density reservoir, 
 namely CV$_1$, and a low density reservoir, CV$_2$, connected by a 
 nanpore.  Flat and repulsive walls surround each entrance of the 
 nanopore. The cylindrical pore in the center has radius $a$ and length $L_{NT}$.
 Right: gray spheres represents solvophobic sites and orange spheres represents 
solvophilic sites in the nanopore.
 From up to down we can see a completely solvophobic nanopore ($\phi = 0.0$), a 
half-solvophilic half-solvophobic 
 nanopore ($\phi = 0.5$) and a completely  solvophilic nanopore ($\phi = 1.0$).}
 \label{fig2}
 \end{figure}

We explore the behavior of these two
types of  fluids confined inside a cylindrical 
 nanopore. In order to
observed the flow, the nanotube is   connected to 
two reservoirs, namely, CV$_1$ on the left of
 the nanopore and CV$_2$ at the right, as depicted 
in the figure~\ref{fig2}. The 
simulation
 box is a parallelepiped with 
dimensions $5L\times L\times L$ in $xyz$ directions,
 with $L = 16$. The tube structure is
 built as a wrapped sheet of spherical particles with 
diameter $\sigma_{\rm NT}=\sigma$. 
 Each particle in the nanotube represents a site that interacts with 
the fluid by an attractive or a repulsive that in
the case of interaction with water are called 
solvophilic or solvophobic interactions. For simplicity
even when interacting with the LJ fluid these attractive
and repulsive interactions will be called solvophilic and solvophobic.
 Solvophobic sites and fluid particles interact through 
 the purely repulsive Weeks-Chandler-Andersen (WCA) 
 potential~\cite{AllenTild} given by
\begin{equation}
\label{LJCS}
U_{ij}^{\rm{WCA}}(r_{ij}) = \left\{ \begin{array}{ll}
U_{{\rm {LJ}}}(r_{ij}) - U_{{\rm{LJ}}}(r_c)\;, \qquad r_{ij} \le r_c\;, \\
0\;, \qquad \qquad \qquad \qquad \quad r_{ij}  > r_c\;,
\end{array} \right.
\end{equation}
where $r_{ij}$ is the distance
between the fluid particle $i$ and the nanopore particle $j$,
$r_c = 2^{1/6}\sigma$ and $\epsilon' = 4.0$. Solvophilic sites and 
fluids interact with the fluid through the strong
attractive (SAT) potential~\cite{Krott14a} namely
\begin{equation}
\label{SAT}
U_{ij}^{\rm{SAT}}(r_{ij}) = \left\{ \begin{array}{ll}
 D_1\left[ \left(\frac{\sigma}{r_{ij}}\right)^{12} - 
 \left(\frac{\sigma}{r_{ij}}\right)^6 \right] + 
 D_2\left[\frac{r_{ij}}{\sigma}\right] - 
 \epsilon_{\rm{SAT}}\;, \qquad r_{ij} \le r_c\;, \\
0\;, \qquad \qquad \qquad \qquad \quad r_{ij}  > r_c\;,
\end{array} \right.
\end{equation}
where $r_{ij}$ is the distance between the fluid particle $i$ and the nanopore particle $j$,
$D_1 = 1.2$, $D_2 = 0.0545$, $\epsilon_{\rm{SAT}} = D_1[(1/r_c)^{12} - 
 (1/r_c)^6] + D_2[r_c]$ and $r_c = 2.0$.
 
Here we explore three cases: systems with solvophobic, solvophilic and 
 mixed nanotube. The three cases are represented by the 
 fraction of solvophilic sites, 
$\phi$, defined as the number of solvophilic 
particles in the nanopore
 structure divided by the total number of particles in the nanopore structure. 
The figure~\ref{fig2} illustrates 
 nanopores fully solvophobic, $\phi = 0.0$, half solvophobic, $\phi = 0.5$, and 
completely solvophilic, $\phi = 1.0$.
 The solvophilic sites are placed starting at the left extrema of the nanopore, 
 in contact with the CV$_1$ reservoir, until the desired fraction $\phi$.
 The nanotube entrances are surrounded by flat walls, which also 
 interact with the fluid particles through the WCA potential. To avoid the flux 
between the CV$_1$ and
 CV$_2$ due the periodic boundary condition, flat walls are placed in the extrema 
of the simulation box
 in the $x$ direction. These walls also interact with the fluid particles through 
the WCA potential.
 
 The temperature of the system was fixed in $T=0.5$ by means of the
 Nose-Hoover thermostat with a coupling parameter $Q=2$.
 This temperature ensures that the bulk system is not in the 
 solid state. For simplicity, we assume that the nanopore 
 atoms are motionless during the entire simulation.
 Periodic boundary conditions are applied in the $y$ and $z$ directions.  A
 time step of $\delta t=0.005$ was adopted.

 A steady state flux through the nanopore is obtained by fixing a higher
 density in the reservoir CV$_1$, $\rho_1 = 0.065$, and a lower
 density
 in CV$_2$, $\rho_2 \approx 10^{-3}$. 
  The initial state is generated using a standard 
 Grand Canonical Monte Carlo (GCMC)
 simulation inside each reservoir, during $5\times10^5$ steps, with the
 initial velocity for each particle obtained from a Maxwell-Boltzmann
 distribution at the desired temperature. The densities inside the reservoirs
 are maintained constant using the Dual Control Volume Grand Canonical
 Molecular Dynamics (DCV-GCMD) method~\cite{DCVGCMD94, Bordin12a}.
 In this method, the desired densities are restored in CV$_1$ and CV$_2$ by
 intercalating the MD steps with a number of GCMC steps inside the
 corresponding control volumes depicted in the figure~\ref{fig2}. In our
 simulations, an initial $5\times10^5$ MD steps were used for
 equilibration of the system, and 50 GCMC steps were performed for
 every 150 MD steps during the DCV-GCMD process. These rate for the
 simulation steps ensures that the density in each reservoir
 changes less than 0.5\%. Final results were
 obtained after performing a total of $5\times10^6$ MD steps
 for each simulation, and averaging over $5$ to $10$ independent runs to
 evaluate relevant physical quantities. Error bars are not shown
 since they are smaller than the data point.
 We have studied the flow induced only by the density gradient
 and the flow induced by the density gradient plus a constant, 
 external gravitational-like force to verify the effects of
 external fields on the flux.

 The axial flux of particles through the nanopore, $J_{x,\rm tube}$, is
 computed by counting the number of particles that cross the channel
 from left to right, $n_{\rm ltr}$, and the particles flowing from right to
 left, $n_{\rm rtl}$~\cite{DCVGCMD94},
 \begin{equation}
  J_{x,\rm pore} = \frac{n_{\rm ltr}-n_{\rm rtl}}{A_{\rm NT}N_{\rm steps} \delta t}\;,
 \end{equation}
 where $A_{\rm NT}=\pi a_{eff}^2$, with $a_{eff} = a - \sigma$ the effective
 radius available for the fluid, $N_{\rm steps}$ is the total number of
 steps used in the simulation, and $\delta t$ is the MD time step. In
 an entirely similar way, we evaluate the flux in the $x$-direction for
 the non-confined case, $J_{x,\rm bulk}$, but now using $A = L\times L$,
 since there is no nanopore in the system. The $J_{x,\rm bulk}$ was used
 to normalize $J_{x,\rm pore}$ in order to compare the bulk and confined flow.
  Here we assume that the fluid-fluid interaction for anomalous and non anomalous
 fluids, Equations~(\ref{Potential}) and (\ref{Potential2}),
 has a cutoff radius $r_{\rm cut} = 3.5$. Two
 nanopore radii where simulated: $a = 1.75$ and $a = 5.0$.
 In all simulations, the tube length is fixed to  $L_{NT}=16$.

 \section{Results and discussion}

\subsection{Water-like Fluid}
  
 \begin{figure}[ht]
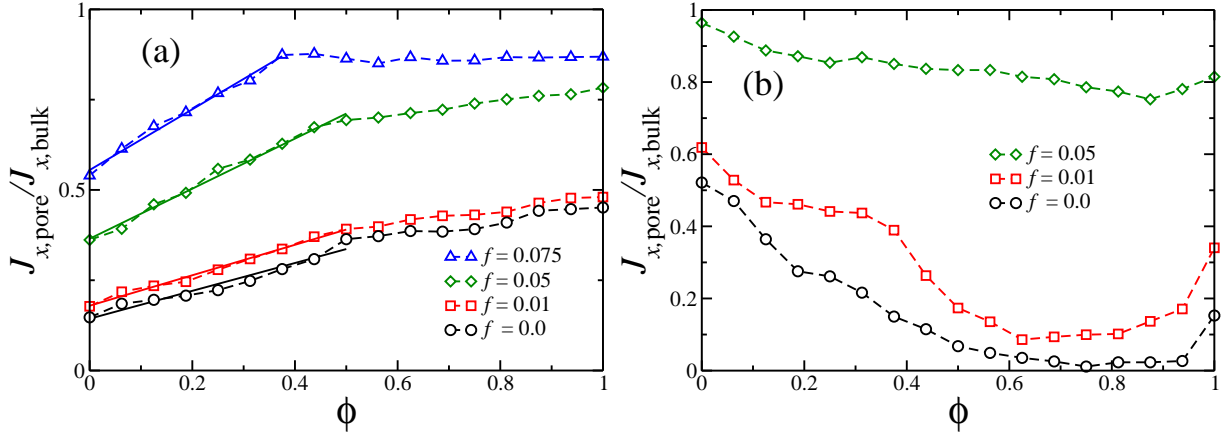

 \begin{center}
 \includegraphics[width=8cm]{fig3a.eps}
  \includegraphics[width=8cm]{fig3b.eps}
   \end{center}
 \caption{Flux of anomalous particles through the 
nanopore, $J_{x,\rm pore}$, in units of 
the non-confined flux, 
 $J_{x,\rm bulk}$, as function of the fraction of solvophilic 
sites $\phi$ 
 for different external forces $f$ and for cylindrical nanopores with radius
 (a) $a = 1.75$ and (b) $a = 5.00$.}
 \label{fig3}
 \end{figure}
 
 Let us first exam what happens with the water-like fluid
confined.
 The figure~\ref{fig3} illustrates the dependence of the flux of
the fluid  with the
 fraction of solvophilic sites, $\phi$, for two values of the
nanotube diameter, $a$, and distinct external forces, $f$. 
The fluid flux is strongly affected by the nanopore radius.
 For the narrowest tube, $a=1.75$, showed in the figure~\ref{fig3}(a),
 the flow increases with  $\phi$ linearly
 until  $\phi\approx 0.5$, and then saturates at 
high values of $\phi$. A linear fit is shown for $\phi < 0.5$. 
For the highest value of $f$  the flow has 
a  maximum at  $\phi\approx 0.4$.
 A similar behavior was observed by Moskowitz et al.~\cite{Mosko14} in
 simulations 
of TIP3P
  water inside narrow nanotubes
 with tunable hydrophobicity.

For the larger tubes the behavior is distinct.
For nanopores with $a=5.0$, shown in the figure~\ref{fig3}(b), the flow
follows what was observed by Goldsmith and Martens~\cite{Martens08}
for TIP3P water confined in nanopores.
The water-like fluid confined in pure
solvophobic pore has a larger flow
when compared with the pure solvophilic pore. However,
this behavior is not linear with $\phi$ but has a minimum.
In order to understand the difference between the 
behavior at small and large $a$ a more detail analysis of
the structure of the liquid inside the tube was performed.
 
 Previous studies have showed that the dynamical
 and structural properties of confined anomalous fluids
 are strongly correlated~\cite{Bordin13a, Krott13b}. 
 Simulations with a high density for the fluid in the source reservoir 
and a similar
TLS potential
 shows that in the narrowest nanopore, $a=1.75$, the fluid is structured 
as a single file. As the nanotube radius is increased, more layers 
are  observed, until the limit
 of large diameters and a bulk-like behavior~\cite{Bordin13a}.
 In order to understand 
the distinct flow patterns with  different nanopore
 radius the density and the fluid
 structure inside each nanopore were analyzed. The density
 profile was evaluated in 
the axial direction $x$
 and in the radial direction, $r_{\rho} = (y^2+z^2)^{1/2}$, and then 
normalized accordingly 
 with
 $$
 \rho_{\rm normalized}(\zeta) = \frac{\rho(\zeta)}{\int \rho(\zeta) d\zeta}\;,
 $$
\noindent with $\zeta = x$ or $\zeta = r_{\rho}$ for axial and radial 
density profile,
respectively.

 \begin{figure}[ht]
 \begin{center}
 \includegraphics[width=8cm]{fig4a.eps}
  \includegraphics[width=8cm]{fig4b.eps}
   \includegraphics[width=8cm]{fig4c.eps}
      \includegraphics[width=8cm]{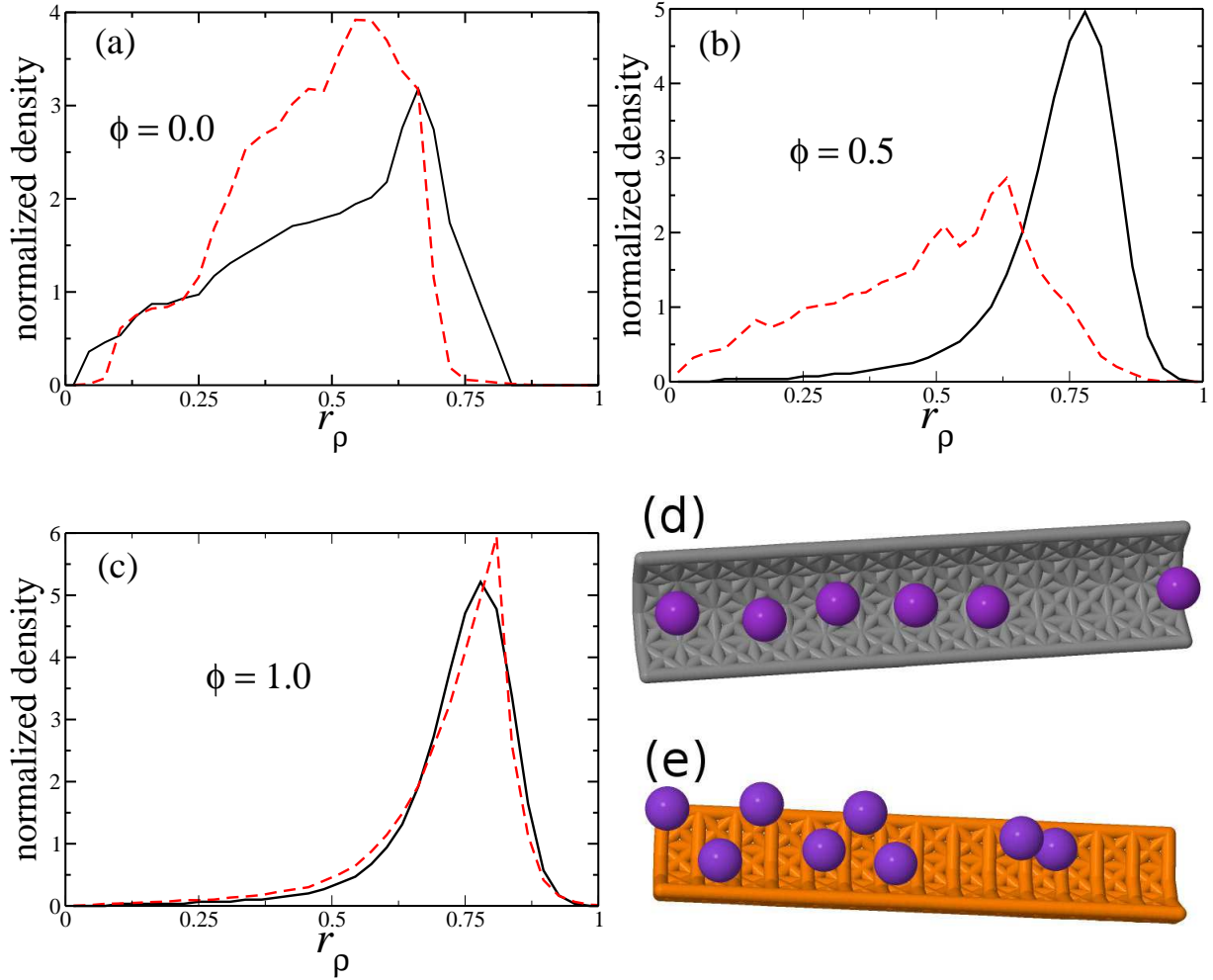}
     \end{center}
 \caption{Normalized density profile inside the nanopore with 
radius $a=1.75$ as function of the radial distance $r_{\rho}$ 
 near the entrance (solid black line) and near the exit (dashed red line)
 of the nanopore cavity, for
 nanopores with (a) $\phi = 0.0$, (b) $\phi = 0.5$ and (c) $\phi = 1.0$. Snapshots
 showing the fluid inside the nanopore 
 for the purely solvophobic (d) and purely Solvophilic (e) cases.}
 \label{fig4}
 \end{figure}
 
The radial density profile
 near the entrance and  near the exit of the nanopore 
illustrates the differences between the flow
for $a=1.75$ and for  $a=5$. The 
figure~\ref{fig4} shows that for functionalized
 pores with $a=1.75$ and  (a)$\phi=0.0$, (b)$\phi=0.5$  and (c)$\phi=1.0$
 the fluid is structured in a single layer. For $\phi=0.0$,  the  
solvophobic nanopore,
 the structure of the fluid 
 at the entrance and at the exit are the same, with the
 fluid occupying a wide range of
 radial positions. For purely solvophilic nanopores, $\phi=1.0$, the 
structure is also the same at the entrance and at the exit. The
fluid, however, assumes a narrow distribution actually
forming a single line  as shown in the figure~\ref{fig4}(c).
 For nanopores partially solvophilic and partially 
solvophobic, as $\phi=0.5$ shown in the figure~\ref{fig4}(b),
 two distinct behaviors were obtained. Near the entrance, where the 
solvophilic sites are located, the fluid forms a narrow single line, 
 while at the exit, where the nanopore is solvophobic, the fluid 
occupies a wider range of positions in the
 radial direction. 
 To clarify this structures, we show in figure~\ref{fig4}(d) the 
 snapshot for the purely solvophobic nanopore and in 
 figure~\ref{fig4}(e) for the purely solvophilic case. 
The narrow distribution of positions of this single line
allows for the huge flow observed in solvophilic walls.

 \begin{figure}[ht]
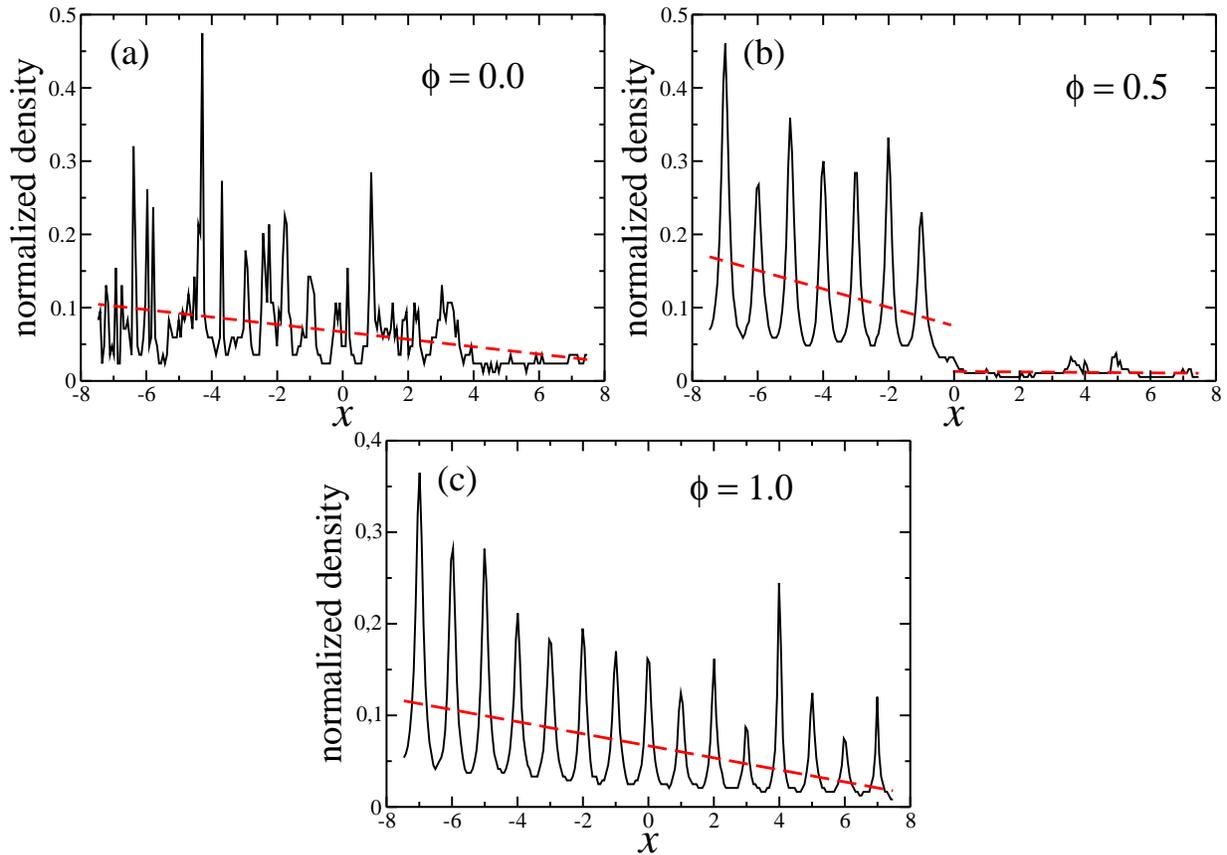

 \begin{center}
 \includegraphics[width=8cm]{fig5a.eps}
  \includegraphics[width=8cm]{fig5b.eps}
   \includegraphics[width=8cm]{fig5c.eps}
     \end{center}
 \caption{Normalized density profile inside the nanopore as function 
of the axial direction $x$, 
 for nanopores with radius $a=1.75$ (a) $\phi = 0.0$, (b) 
$\phi = 0.5$ and (c) $\phi = 1.0$.
 The dashed red lines are the linear fits.}
 \label{fig5}
 \end{figure}

In the figure~\ref{fig5}(a) the normalized density profile in the axial 
direction for narrow
 solvophobic nanopores is illustrated. The profile indicates a fluid without 
structure in the $x$-direction.
 The combination of the
 figure~\ref{fig5}(a) with the figure~\ref{fig4}(a) indicates
that the fluid is in a low density fluid
 phase. The dashed line in the figure~\ref{fig5}(a) is a 
linear fit from the curve. For solvophobic walls fewer particles
enter in the tube and their flow is not equally spaced. 
 The decay indicates a density gradient inside the nanopore.
 
The solvophilic case, 
shown in the figure~\ref{fig5}(c), the fluid 
 exhibits a layered structure in the $x$-direction due to 
the interaction with the wall. 
 Each peak in the figure~\ref{fig5}(a) corresponds to the particles 
in the snapshot, Figure~\ref{fig4}(e).
 Two  length scales fluids tend to assume structures in the first 
or in the second length scale.
 In the confined case, the distance between the layers is approximately 
 the particle radius. This is the first length scale in the potential
 defined by the Eq.~\ref{Potential}.
 The linear fit in the $\rho(x)$ curve also shows the existence of
 a density gradient in the nanopore. Then, in a solvophilic
 nanopore the fluid particles are arranged in the zig-zag single 
file layered structure,
 moving from one preferable distance to another. This
more rigid structure in zig-zag moves faster
than the disordered structure observed in 
the case of the solvophobic confinement.

In a partially solvophilic, 
 partially solvophobic nanopore the two distinct structural behaviors
 in the $x$-direction are observed. The solvophilic region 
is a high density, well structured
 fluid, while in the solvophobic region there is a low density, non 
structured fluid,
 as the figure~\ref{fig5}(c) shows. The linear fit in the structured region
 shows a high slope. Comparing with the Fick's Law for a one dimensional flow,
 $ J = -D (d\rho/dx)\;,$,
 where D is the diffusion coefficient, we can see that a higher slope
 will lead to a higher flow, which can easily leak through the solvophobic
 region. Therefore, the mixture of solvophilic and solvophobic
 sites plays a important role in the flow. In this way, the fluid flow 
 increases when solvophilic  sites are added to the nanopore until the threshold. This 
picture is quite similar to the description of 
Moskowitz et al~\cite{Mosko14} and Hummer et al~\cite{Hummer01}.

 \begin{figure}[ht]
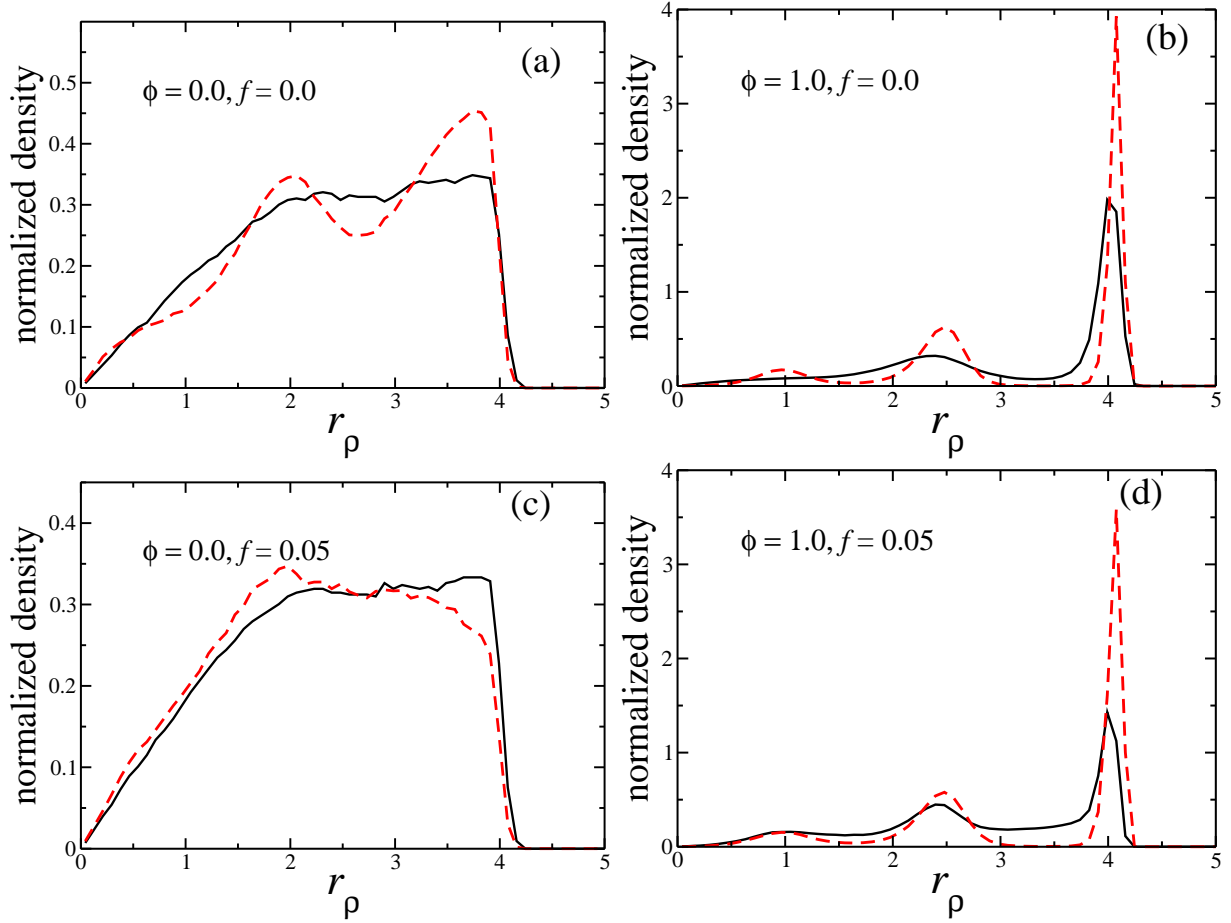

 \begin{center}
 \includegraphics[width=8cm]{fig6a.eps}
  \includegraphics[width=8cm]{fig6b.eps}
   \includegraphics[width=8cm]{fig6c.eps}
   \includegraphics[width=8cm]{fig6d.eps}
     \end{center}
 \caption{Normalized density profile inside the nanopore with $a=5.0$  as function of 
the radial distance $r_{\rho}$
 near the entrance (solid black line) and near the exit (dashed red line) of the 
nanopore cavity, for
 nanopores with (a) $\phi = 0.0$ and $f = 0.0$, (b) $\phi = 1.0$ and $f = 0.0$, (c) 
$\phi = 0.0$ and $f = 0.05$
 and (d)$\phi = 1.0$ and $f = 0.05$.}
 \label{fig6}
 \end{figure}

\begin{figure}[ht]
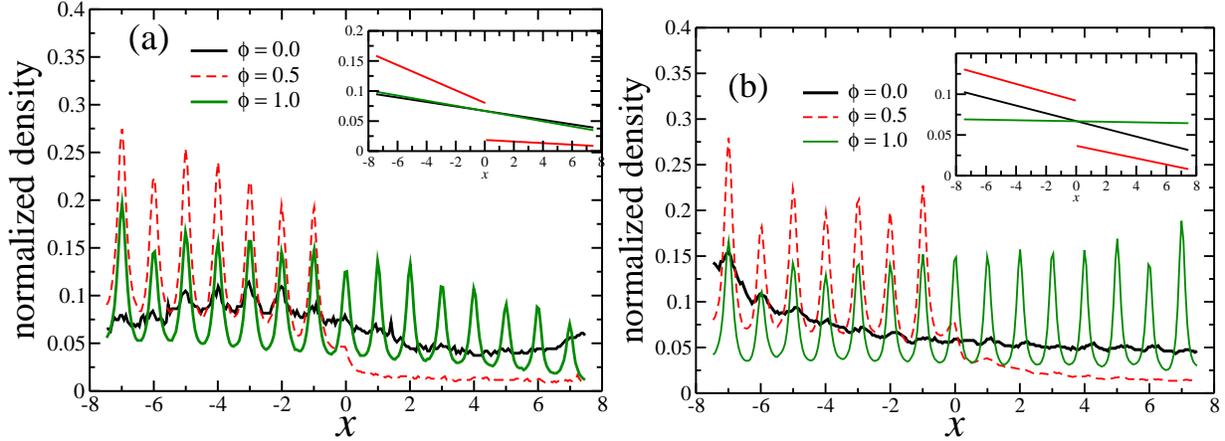

 \begin{center}
 \includegraphics[width=8cm]{fig7a.eps}
  \includegraphics[width=8cm]{fig7b.eps}
     \end{center}
 \caption{Normalized density profile inside the nanopore as function of the axial direction $x$, 
 for nanopores with radius $a=5.0$ (a) $f = 0.0$ and (b) $f = 0.05$.}
 \label{fig7}
 \end{figure}

Now, let us exam how is  the density profile
 near the entrance and  near the exit of the nanopore 
 for  $a=5$.
The figures~\ref{fig6}(a) and (c) show the normalized
density as a function of the radial distance in the 
case of purely solvophobic nanotube for  the external forces
$f=0.0$ and $f=0.05$, respectively. In both cases the fluid
does not show a well defined structure, with a fluid-like profile. 
No layering is observed and therefore not friction
between the layers is observed. The figures ~\ref{fig6}(b) and (d) show
the radial density profile
for purely solvophilic nanopores for $f=0.0$ and  $f=0.05$ respectively. Both 
near the entrance and at the end of the nanotube layering 
is observed with the contact layering showing 
a high density that increases at the end of the tube.
The distance between layers corresponds to the second
length scale what suggests that the system 
becomes quite stable in this configuration, making
hard for the particles to move. This 
results is confirmed by the normalized axial
density as a function of $x$ illustrated in the 
 figure~\ref{fig7}. For the pure solvophobic system 
the density is uniform not indicating any organization, liquid-like.
For the pure solvophilic, the system form a periodic
structure what suggests almost a solid-like system~\cite{Li12,Martens08}.

Then we can understand how the results of 
indicating that flow for solvophobic walls are 
faster~\cite{Li12,Martens08} are
not contradictory to the results
indicating that the flow for
solvophilic walls are more rapid~\cite{Mosko14,Hummer01}.
In the first case~\cite{Li12,Martens08} the system well defined
layers with the fluid wall particles being fully commensurated, in the
second case either just a single layer is formed or
the flow is structured because wall-fluid are not 
commensurated~\cite{Mosko14,Hummer01}.

\subsection{Lennard-Jones-like  Fluid}

 \begin{figure}[ht]
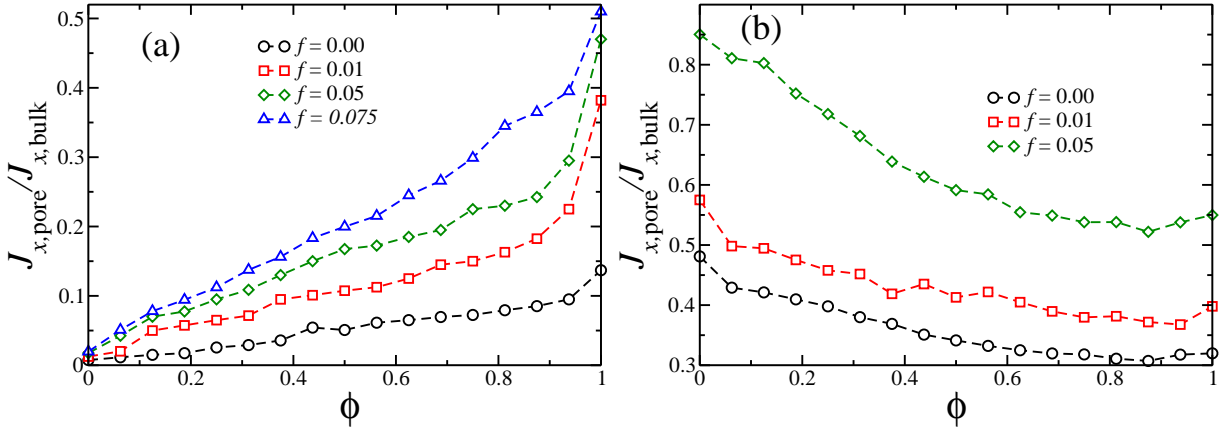

 \begin{center}
 \includegraphics[width=8cm]{fig8a.eps}
  \includegraphics[width=8cm]{fig8b.eps}
   \end{center}
 \caption{Flux of non anomalous particles through the nanopore, $J_{x,\rm pore}$, in units of 
the non-confined flux, 
 $J_{x,\rm bulk}$, as function of the fraction of solvophilic sites $\phi$ 
 for different external forces $f$ and for cylindrical nanopores with radius
 (a) $a = 1.75$ and (b) $a = 5.00$.}
 \label{fig8}
 \end{figure}

Next, we analyze if the flow in the system in which only one length scale 
is present differs from the flow in the two length
scale system what could explain the difference 
between the flow of water and the flow of gases.

 \begin{figure}[ht]
 \begin{center}
 \includegraphics[width=8cm]{fig9a.eps}
  \includegraphics[width=8cm]{fig9b.eps}
   \includegraphics[width=8cm]{fig9c.eps}
   \includegraphics[width=8cm]{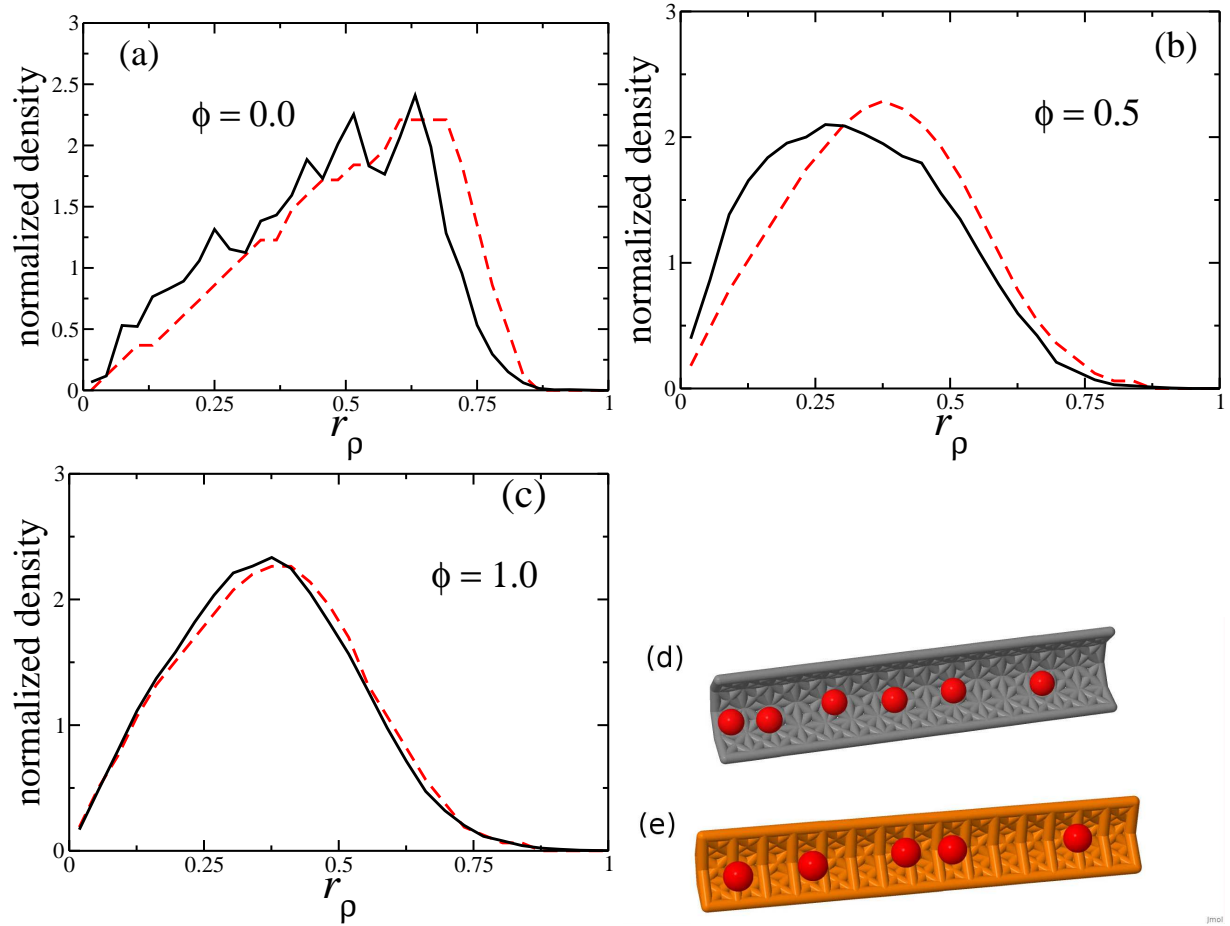}
     \end{center}
 \caption{Normalized density profile for the non anomalous fluid confined in a nanopore with 
radius $a=1.75$ as function of the radial distance $r_{\rho}$ 
 near the entrance (solid black line) and near the exit (dashed red line)
 of the nanopore cavity, for
 nanopores with (a) $\phi = 0.0$, (b) $\phi = 0.5$ and (c) $\phi = 1.0$.
 Snapshots for the solvophobic (d) and solvophilic (e) cases.}
 \label{fig9}
 \end{figure}

\begin{figure}[ht]
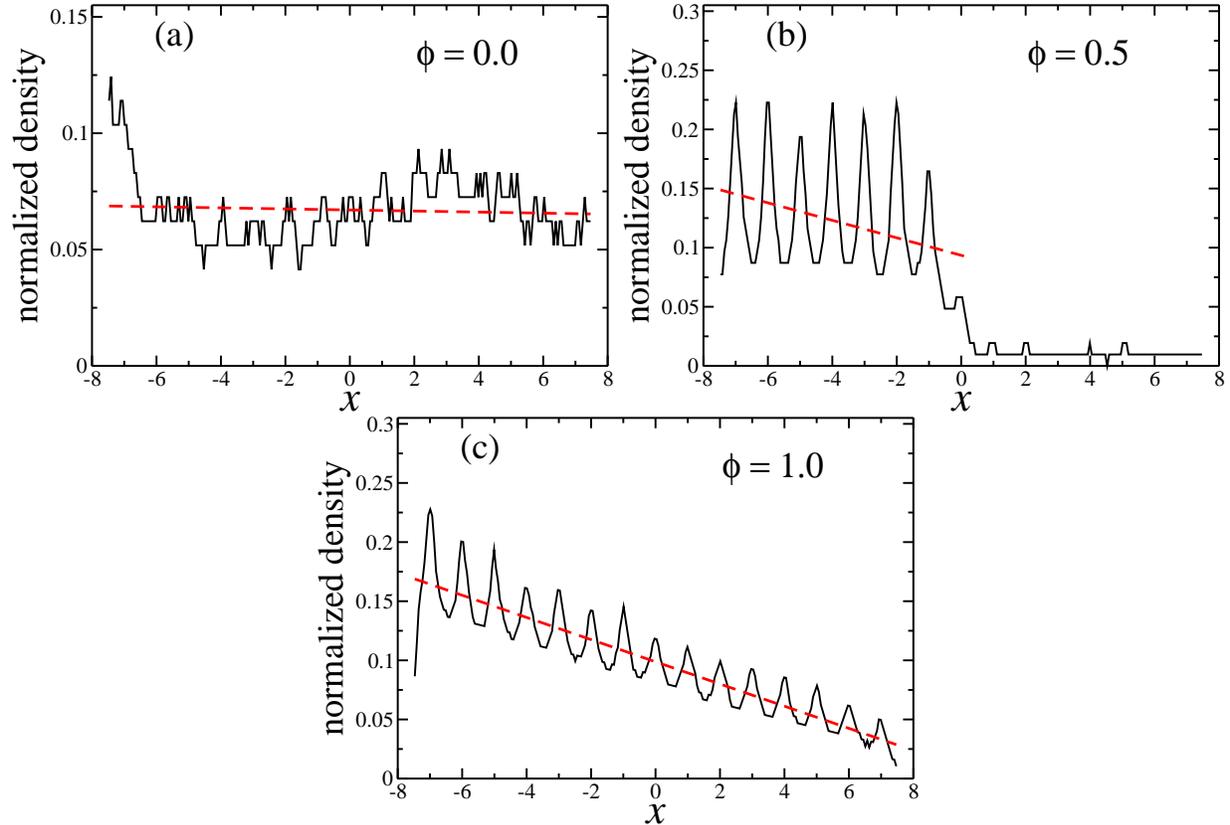

 \begin{center}
 \includegraphics[width=8cm]{fig10a.eps}
  \includegraphics[width=8cm]{fig10b.eps}
   \includegraphics[width=8cm]{fig10c.eps}
     \end{center}
 \caption{Normalized density profile inside the nanopore as function 
of the axial direction $x$, 
 for nanopores with radius $a=1.75$ (a) $\phi = 0.0$, (b) 
$\phi = 0.5$ and (c) $\phi = 1.0$.
 The dashed red lines are the linear fits.}
 \label{fig10}
 \end{figure}

The figures~\ref{fig8}(a) and (b) illustrates the normalized flow for
the LJ fluid for (a) $a = 1.75$ and (b) $a = 5$ respectively.
The comparison of  the figure~\ref{fig3} with the
figure~\ref{fig8} shows that for both TLS and OLS potentials the 
flow is larger for the solvophilic when compared
with the hydrophobic confinement. The mechanism
for the increase of the flow with solvophilicity is the
increase in density as the solvophilic confinement allow for 
more particles to enter into the tube~\cite{Hummer01}.
This line, however, as illustrated in the 
 figure~\ref{fig10} is not the same for the different types
of confinement. For the solvophilic confinement the line is 
more organized and the particles in this single line move 
in jumps from one structure to the other. Since one line
is present this movement shows almost no friction
and the particles move faster in the organized structure (solvophilic) 
than in the disorganized (solvophobic) arrangement.

The difference between the water-like and 
the LJ systems is the amount of flow
that is higher for the TLS potential in the 
case of very small confinement, namely $a=1.75$. This result that
is  consistent with experimental observations~\cite{Holt06,Qin11} 
 can be understood as follows is observed comparing the
figure~\ref{fig3}(a) with the figure~\ref{fig8}(a). In the case
of the TLS potential the single line is more compact as
can be seen by comparing  the 
figures~\ref{fig4} and ~\ref{fig5}  with the figures~\ref{fig9} and
~\ref{fig10} since the two length
scales allows for more mobility. 

\begin{figure}[ht]
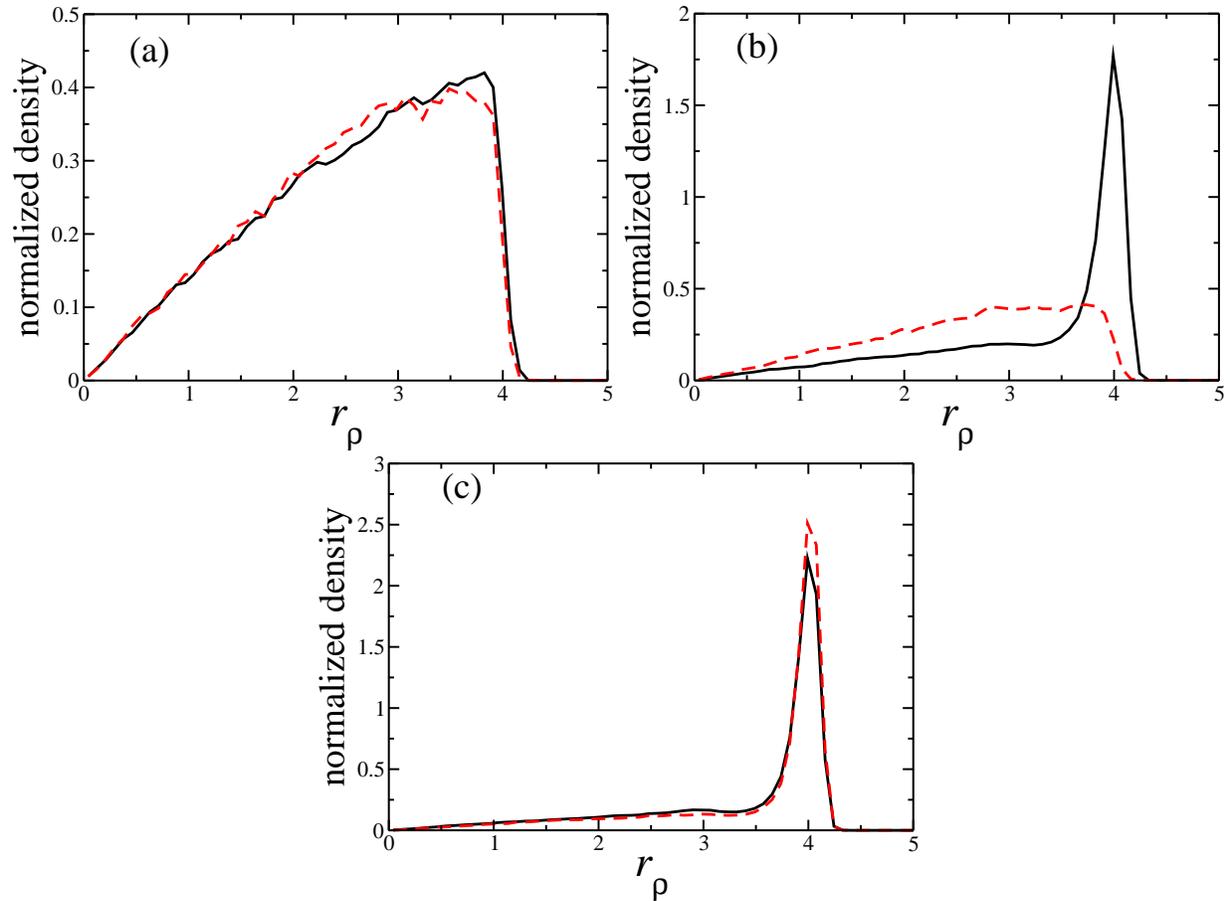

 \begin{center}
 \includegraphics[width=8cm]{fig11a.eps}
  \includegraphics[width=8cm]{fig11b.eps}
   \includegraphics[width=8cm]{fig11c.eps}
        \end{center}
 \caption{Normalized density profile inside the nanopore with $a=5.0$  as function of 
the radial distance $r_{\rho}$
 near the entrance (solid black line) and near the exit (dashed red line) of the 
nanopore cavity, for}
 \label{fig11}
 \end{figure}

The difference between the TLS and OLS is also observed
for $a = 5$ case. While for the TLS the system
forms layers, in the OLS no layer is observed as shown
in the figure~\ref{fig11}. 
This difference, however, does not impact the 
decrease in flux as the system becomes more 
solvophilic. This as in the case of the TLS potential
results from our choice of particles that a commensurated
with the wall. Then, as the system becomes
more solvophilic the fluids organize on the wall 
forming a layer that increases friction.

 \section{Conclusion}
 
 In this paper we addressed two questions related 
to the flow of fluids under nanoconfinement. First, 
we showed that the flow increase related to the change
from solvophobic to solvophilic only happens 
if the particles and wall are not commensurated or
if the flow occurs in a single line, what favors decorrelation
with the wall. This effect is observed both 
for water-line and LJ type of flow.

Then, we compared the flow of water-like with 
LJ line fluid in order o understand the reason behind
the faster flow of water when compared with
non anomalous fluids as gases. Our results indicates
that water-like fluids employ the two length scales to
form more compact structures decreasing friction 
and allowing for a faster mobility.

 \section{Acknowledgments}
 We acknowledge financial support from the Brazilian Agencies CNPq
 and FAPERGS.


\end{document}